\documentclass{svproc}

\usepackage{url}
\usepackage{graphicx}%
\usepackage{multirow}%
\usepackage{amsmath,amssymb,amsfonts}%
\usepackage{mathrsfs}%
\usepackage[title]{appendix}%
\usepackage{xcolor}%
\usepackage{textcomp}%
\usepackage{manyfoot}%
\usepackage{booktabs}%
\usepackage{algorithm}%
\usepackage{algorithmicx}%
\usepackage{algpseudocode}%
\usepackage{listings}%

\usepackage{subfig}
\usepackage[labelfont=bf]{caption}
\usepackage{rotating}
\usepackage{nicematrix}
\usepackage{makecell}
\usepackage{xcolor}
\usepackage{hyperref}
\usepackage[separate-uncertainty = true,multi-part-units=single]{siunitx}
\begin{document}
\mainmatter
\title{Soft Tissue Simulation and Force Estimation from Heterogeneous Structures using Equivariant Graph Neural Networks}
\titlerunning{cEGCL}
\author{Madina Kojanazarova \and Sidaty El Hadramy \and Jack Wilkie \and Georg Rauter \and Philippe C. Cattin}
\authorrunning{Madina Kojanazarova et al.} 
%
\tocauthor{Madina Kojanazarova, Sidaty El Hadramy, Jack Wilkie, Georg Rauter, Philippe C. Cattinn}
\institute{University of Basel, Department of Biomedical Engineering, Allschwil, Switzerland\\
\email{M.Kojanazarova@unibas.ch}
}

\maketitle

\begin{abstract}
Accurately simulating soft tissue deformation is crucial for surgical training, pre-operative planning, and real-time haptic feedback systems. 
While physics-based models such as the finite element method (FEM) provide high-fidelity results, they are often computationally expensive and require extensive preprocessing. 
We propose a graph neural network (GNN) architecture that predicts both tissue surface deformation and applied force from sparse point clouds.
The model incorporates internal anatomical information through binary tissue profiles beneath each point and leverages E(n)-equivariant message passing to improve robustness.
We collected experimental data that comprises a real silicone and bone-like phantom, and complemented it with synthetic simulations generated using FEM. 
Our model achieves a comparable performance to a baseline GNN on standard test cases and significantly outperforms it in rotated and cross-resolution scenarios, showing a strong generalization to unseen orientations and point densities. 
It also achieves a significant speed improvement, offering a solution for real-time applications. 
When fine-tuned on experimental data, the model maintains sub-millimeter deformation accuracy despite limited sample size and measurement noise.
The results demonstrate that our approach offers an efficient, data-driven alternative to traditional simulations, capable of generalizing across anatomical configurations and supporting interactive surgical environments. 

\end{abstract}
\section{Introduction}
Simulating the behavior of biological tissue offers significant benefits across various areas of healthcare \cite{agha2015,costello2022,Cotin1999-sw}. 
In surgical training, realistic tissue models provide a safe environment for surgeons to practice procedures, thereby enhancing their skills and confidence \cite{al-kadi2012,meling2021,Tai2018}. 
For pre-operative planning, such simulations enable clinicians to anticipate potential complications by predicting how tissues will respond during surgery. 
In surgical guidance, especially when integrated with augmented reality, real‑time simulations of tissue deformations are used intraoperatively to update the visualization of internal structures, which can greatly improve clinical decision-making and precision \cite{Hadramy2024-dk,Haouchine2013-ts}. 
Furthermore, in medical robotics, particularly in applications involving flexible robots or interactions with soft tissues, simulation of biological structures is crucial for achieving accurate and responsive control of the robotic systems \cite{Pore2021,Zrinscak2023}.

To be effectively integrated into these healthcare applications, simulations of biological tissue must be both \textbf{accurate} and \textbf{real-time} \cite{Mendizabal2020-gs}. 
High accuracy ensures that the mechanical response of the simulated tissue closely mimics real biological behavior, while real-time performance is critical in interactive scenarios such as surgical training, where immediate feedback and responsiveness are required \cite{costello2022}. 
In the context of \textbf{surgical training}, integrating haptic feedback further enhances realism by allowing trainees to physically feel the interaction with biological tissues \cite{abinaya2024}. 
This tactile sensation creates a more immersive and effective learning environment, closely replicating the conditions of actual interventions \cite{Tai2018,Tai2016}. 
However, delivering realistic haptic feedback requires modeling the interaction forces in a way that captures the mechanical behavior of the tissue. This is especially important for conveying differences between anatomical structures and detecting changes in tissue properties during manipulation 
\cite{Coles2011,Castillejos2016}.

Several studies have demonstrated the effectiveness of physics-based methods in simulating the behavior of biological tissues \cite{Haouchine2013-ts,Mazier2024-bx}. 
These simulations typically rely on principles of continuum mechanics and often employ finite element methods (FEM) to solve the numerical systems governing organ and tissue dynamics. 
Although these models can provide high accuracy, they are computationally expensive, often resulting in long calculation times that limit their use in real-time surgical training applications \cite{Nguyen2020}. 
To address this limitation, learning-based surrogate models have been introduced as an effective alternative. 
These models are trained on data generated from simulations and offer a real-time approximation of tissue behavior upon training, making them well-suited for time-sensitive applications. 

Recently, deep neural networks have emerged as powerful surrogate models for simulating tissue deformation. 
Odot \textit{et al.} \cite{Odot2022-sk} demonstrated that a multilayer perceptron (MLP) can learn the deformation behavior of a regular beam. 
Building on this, Mendizabal \textit{et al.} \cite{Mendizabal2020-gs} introduced U-Mesh, a U-Net-based architecture capable of learning the deformation of complex, patient-specific geometries such as the liver. 
This was further extended by introducing HyperU-Mesh \cite{El_Hadramy2025} and Deform Any Liver (DAL) \cite{El_Hadramy2025-tt}, which generalize to varying material properties and anatomical geometries, enabling accurate deformation predictions across a wide range of liver models. 
Another promising class of surrogate models is Graph Neural Networks (GNNs), which are particularly well-suited for modeling deformable structures due to their ability to handle unstructured data such as point clouds and meshes \cite{Salehi2022-wy}. 
Deshpande \textit{et al.} \cite{Deshpande2024-fy} introduced MagNet, a framework that extends traditional convolutional neural networks to operate on arbitrary graph-structured data. 
This allows the model to effectively capture complex spatial relationships and deformations in irregular geometries. 
Kojanazarova \textit{et al.} \cite{cGNN2025} introduced a Conditional Graph Neural Network model that predicts both surface deformation and applied force from sparse poking data. 
While all these approaches are promising, they are either unable to provide estimations of the applied forces required for haptic feedback or are limited to predicting deformations in homogeneous tissues, which restricts their applicability in realistic surgical training scenarios with complex heterogeneous anatomical scenarios.

\textbf{Proposed Method} In this work, we propose a novel pipeline for soft tissue simulation that overcomes these limitations and brings us closer to realistic, interactive surgical training. 
First, we introduce heterogeneous object modeling by incorporating the internal anatomical structure.
Specifically, we include soft tissue over hard (bone-like) embedded tissue in the learning process. 
We collect experimental data from a custom-made phantom consisting of silicone molded around a 3D-printed internal geometry, while simultaneously measuring surface deformation and applied force. 
To overcome the limitations of sparse and noisy experimental data, we generate high-resolution training data using FEM simulations, enabling the model to learn richer deformation patterns.
We then adapt the Equivariant Graph Convolutional Layer \cite{ecgnn} into 
\textit{Conditional Equivariant Graph Convolutional Layer} (cEGCN) in our architecture, which improves inference speed and enables the model to generalize across unseen orientations and poses without explicit rotational data augmentation. 
Furthermore, we demonstrate the pipeline's transferability by applying the model trained on FEM simulations to experimentally collected data. 
Finally, our method predicts both deformation and applied force, facilitating integration into haptic feedback systems for real-time surgical simulation.

\begin{figure}[b!]
    \centering
    \subfloat[]{\includegraphics[width=0.35\textwidth]{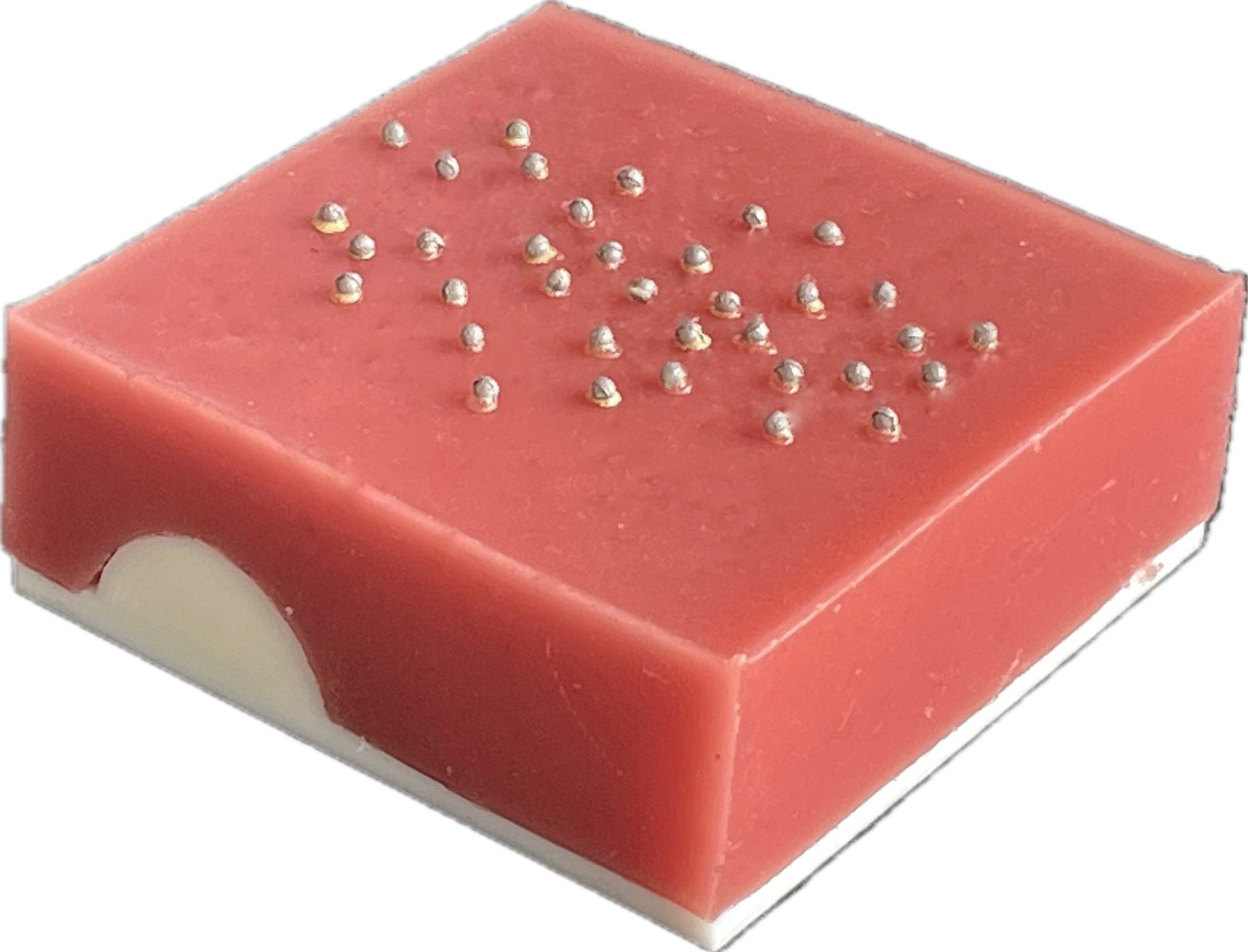}\label{3d_print}}
    \subfloat[]{\includegraphics[width=0.4\textwidth]{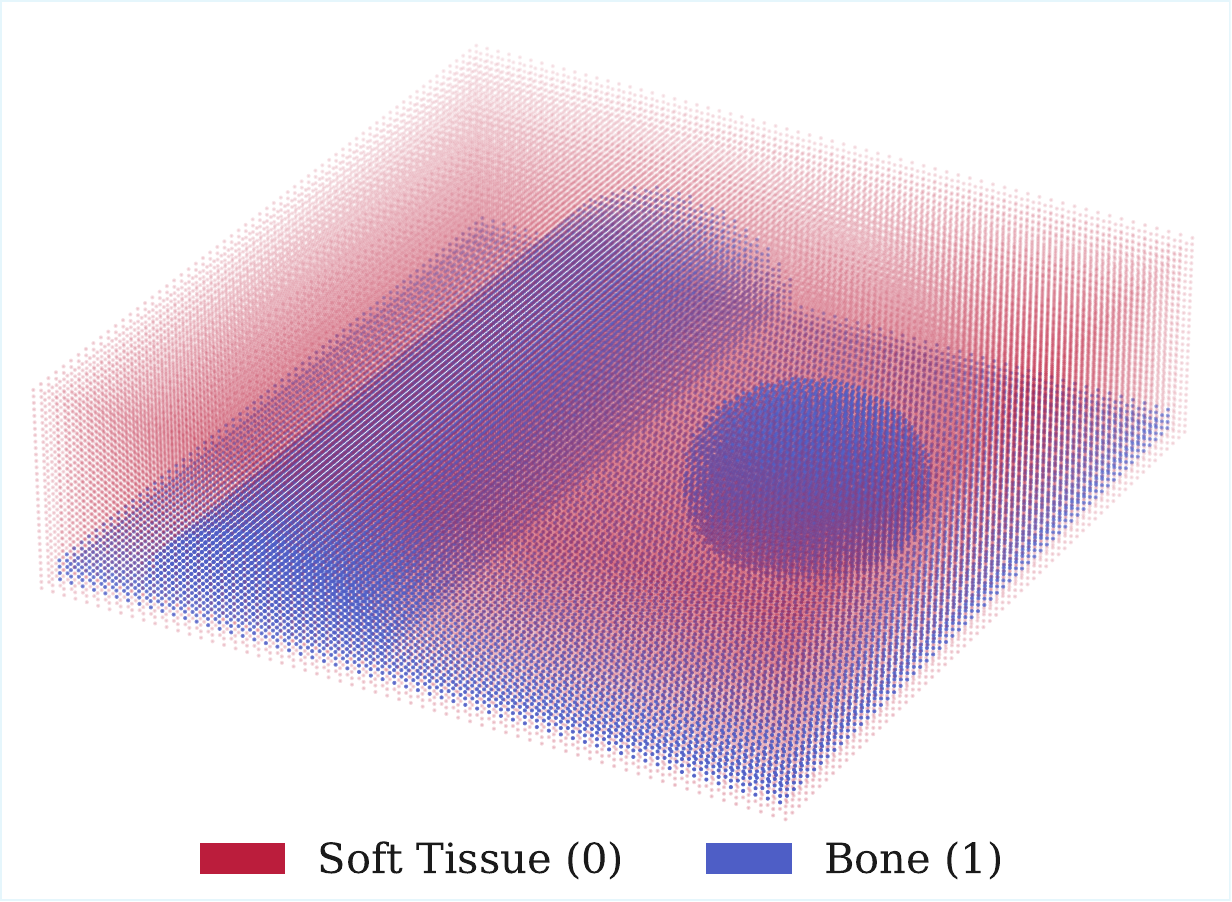}\label{binary}}
    \caption{\textbf{(a)} 3D-printed shape embedded in silicone with glued \qty{3}{\mm} passive reflective markers to record the surface deformation. \textbf{(b)} Binary image volume representing the bone shape and silicone.} 
    \label{volume}
\end{figure}

\section{Methods}\label{sec2}

\subsection{Experimental Setup}
We acquired experimental surface tracking and force data of silicone and a 3D-printed bone structure underneath (Fig.~\ref{3d_print}). 
The bone structure was 3D-printed using fused deposition modeling (FDM) technology with white ABS material on a Stratasys Fortus 250 printer.
It measures \qtyproduct[product-units = power]{100 x 100}{\mm} in width, and the addition of a silicone mixture brings the total height to \qty{32}{\mm}.
The silicone mixture was prepared by mixing the two components of silicone rubber (Eurosil 4 A+B, Schouten SynTec, the Netherlands) with an oil softener S (Eurosil Softener, Schouten SynTec, the Netherlands) in a 1A:1B:2.5S ratio.
The surface tracking was collected using five Qualisys motion capture cameras with passive surface markers (Qualisys AB, Sweden).
The calibration of the Qualisys system reported the average residual error per camera as \qty{0.26}{\mm}, and the calibration with a standard deviation of \qty{0.20}{\mm}, indicating sub-millimeter accuracy of 3D marker reconstruction.
The surface had 40 pieces of \qty{3}{\mm} markers, including the indentation tip with a Nano17 force sensor attached to record the forces applied to the object.
We collected the experimental data by indenting in $39$ locations on the surface and then extracted static deformations at increments of \qty{0.1}{\N}, resulting in $1,315$ static deformation simulations.

\subsection{FEM Simulation}
To generate a larger body of training data, we performed quasi-static finite element simulations in COMSOL Multiphysics \cite{comsol2023}, modeling induced deformation of heterogeneous soft tissue. 
The model included a soft, nearly incompressible hyperelastic region (representing the silicone, with $E = \SI{5}{\kilo\pascal}$ and $\nu = 0.02$) and embedded rigid bone structure. 
Deformations were driven by prescribed displacements up to \qty{30}{\mm} applied to a probe region over \qty{1}{\s}, while boundary conditions constrained the lower surfaces and allowed free motion elsewhere. 
Contact between tissue and bone was modeled using a penalty method with a manually tuned factor $f_p = 175$. 
The mesh consisted of approximately $9,000$ tetrahedral elements with quadratic serendipity shape functions, refined near the contact zone and solved using the PARDISO solver with adaptive time-stepping. 
Each simulation involved $~45,000$ degrees of freedom. 
We sampled $592$ random probe locations on the tissue surface with inclinations up to $41$° from the normal.
To avoid redundancy from small intermediate steps, we retained only those simulation states where the applied force had increased by at least \qty{0.1}{\N} compared to the previously stored state. 
This resulted in $12,387$ unique static deformation configurations with corresponding ground-truth displacement fields and force values, which were used to train a neural network on sparse surface points as input.

\subsection{Tissue Value Extraction}\label{sec:methods_tissue}
To provide the network with information about the internal tissue composition beneath each surface point, we constructed a virtual point-based representation of the experimental sample, including both soft tissue (silicone) and bone. 
This digital volume was assembled by combining the CAD model of the 3D-printed bone with additional uniformly distributed points representing the surrounding silicone, filling a volume of \qtyproduct[product-units = power]{100 x 100 x 32}{\mm} to match the physical dimensions of the sample.
Each point in this volume was assigned a binary label indicating tissue type: 1 for bone and 0 for soft tissue (see Fig.~\ref{binary}).
To extract structural context for each surface point, we sampled along the depth axis (z-direction) at 128 regularly spaced intervals, collecting the tissue type encountered at each step.
This produced a 128-dimensional binary vector per surface point, encoding the vertical material distribution beneath it.
As described in the following sections, these binary profiles were used as point-wise condition features in the model, enabling it to learn deformation behavior in relation to underlying structural heterogeneity.

\subsection{Model Architecture}
We present a graph neural network model designed to predict both the surface displacement and the applied force, given an input point cloud with tissue-specific features and conditions that include the location of the applied force.
The model is based on the Equivariant Graph Convolutional Layer (EGCL) introduced in \cite{ecgnn}, which we adapted into a custom message-passing layer called \textit{Conditional Equivariant Graph Convolution Layer} (cEGCL). 

\subsubsection{Conditional Equivariant Graph Convolution Layer}
We retain the core update rules for node features and coordinates (see Eq.~(5) and Eq.~(6) in \cite{ecgnn}), but adapt Eq.~(3) and Eq.~(4) to our application by introducing external condition inputs and simplifying the edge structure.

Specifically, we remove the edge attributes $a_{ij}$ from Eq.~(3) in \cite{ecgnn} and instead construct graph edges dynamically using $k$-nearest neighbors. 
We also extend the position update function Eq.~(4) in \cite{ecgnn}, to incorporate information about the applied force location. Our message-passing is defined as:

\begin{equation}
\mathbf{m}_{ij} = \phi_e\left(\mathbf{h^\textit{l}}_i, \mathbf{h^\textit{l}}_j, \left\| \mathbf{x^\textit{l}}_i - \mathbf{x^\textit{l}}_j \right\|^2\right)
\end{equation}

\begin{equation}
\mathbf{x^{\textit{l}+1}}_i = \mathbf{x^\textit{l}}_i + \phi_{cs}(\mathbf{x^\textit{l}}_i - \mathbf{C}_s) + \phi_{ce}( \mathbf{x^\textit{l}}_i - \mathbf{C}_e) + \frac{1}{k-1} \sum_{j \neq i} (\mathbf{x^\textit{l}}_i - \mathbf{x^\textit{l}}_j) \cdot \phi_x(\mathbf{m}_{ij})
\end{equation}

Here, $\phi_e$, $\phi_x$, $\phi_{cs}$, and $\phi_{ce}$ are multilayer perceptrons (MLPs), and \( \frac{1}{k-1} \) is a normalization factor applied over the \( k-1 \) neighbors (excluding the point itself). The condition vectors \( \mathbf{C}_s \) and \( \mathbf{C}_e \) represent the start and end points of the applied force on the surface. By incorporating them into the coordinate update, the model becomes aware of where interaction occurs, enabling spatially localized and context-aware deformation prediction.

\begin{figure}
    \centering
    \includegraphics[width=\linewidth]{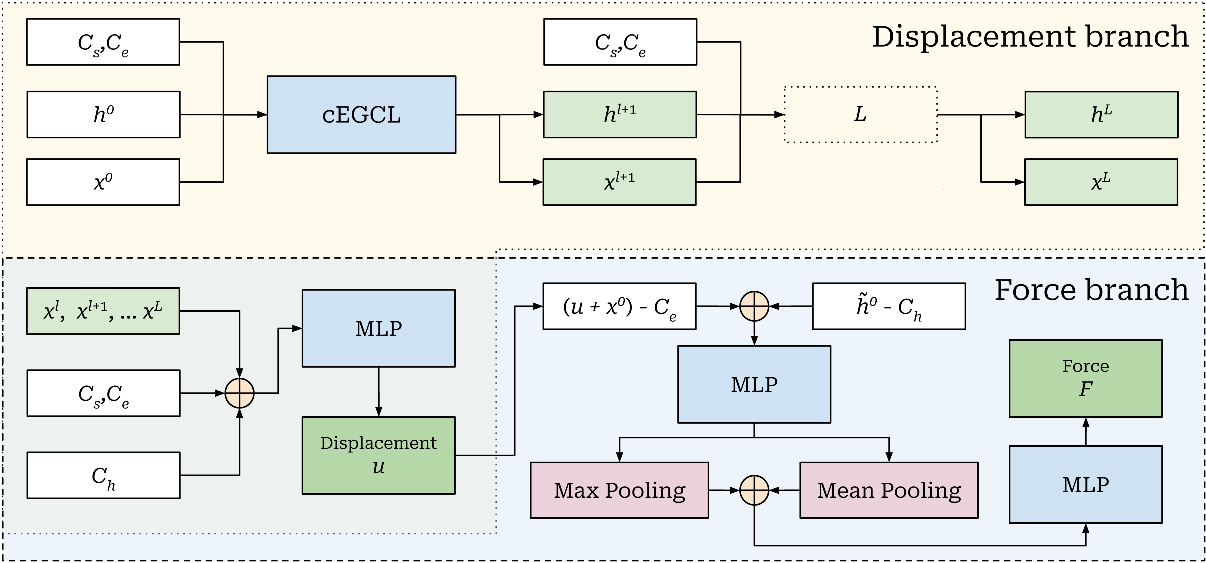}
    \caption{Proposed graph neural network architecture. The input is a surface point cloud $\mathbf{x^0} \in \mathbb{R}^{N \times d}$ with point-wise tissue features $\mathbf{h^0} \in \mathbb{R}^{N \times d_h}$. Conditions $\mathbf{C}_s, \mathbf{C}_e$ indicate the start and end of the applied force location. The Conditional Equivariant Graph Convolution Layers (cEGCLs) update coordinates $\mathbf{x}^l$ and features $\mathbf{h}^l$ across $L$ layers. Outputs from all layers, along with conditions $\mathbf{C}_s, \mathbf{C}_e$ and their tissue features $\mathbf{C}_h$, are concatenated ($\oplus$) and used by two branches, displacement branch: an MLP predicts per-point displacement $\mathbf{u} \in \mathbb{R}^{N \times d}$, and force branch that predicts the scalar force $F \in \mathbb{R}$.}
    \label{cEGNN}
\end{figure}

Having defined the core message-passing operation in the form of the cEGCL, we now describe how it is integrated into the overall model architecture.
The architecture, shown in Figure~\ref{cEGNN}, consists of two main components: (i) a displacement prediction branch and (ii) a force prediction branch.

\subsubsection{Displacement Prediction}
The displacement branch is composed of $L$ stacked cEGCL layers. 
Each layer operates on a point cloud \( \mathbf{x}^l \in \mathbb{R}^{N \times d} \), where \( N \) is the number of surface points and \( d = 3 \) corresponds to their 3D coordinates. 
In addition, each point has associated features \( \mathbf{h}^l \in \mathbb{R}^{N \times d_{h}} \), where \( d_{h} = 256 \) encodes a flattened vector representing the binary tissue values (soft or hard) and their corresponding z-axis positions sampled beneath each point, as described in Section~\ref{sec:methods_tissue}.
The network also takes a condition vector $\mathbf{C}_s, \mathbf{C}_e \in \mathbb{R}^3$ representing the 3D start and end coordinates of the applied force (i.e., the probe's contact line on the surface). 
In addition, we define a condition feature vector \( \mathbf{C}_h \in \mathbb{R}^{4 \times 128} \), which encodes the binary tissue profile and corresponding 3D coordinates sampled along the straight line from \( \mathbf{C}_s \) to \( \mathbf{C}_e \), then flattened into a single vector (Fig.~\ref{cond}). 
This provides spatial context along the direction of interaction.
Each cEGCL layer updates both the point coordinates and their features, producing outputs \( \mathbf{x}^{l+1} \) and \( \mathbf{h}^{l+1} \).

After $L$ layers, the outputs from all layers \( \mathbf{x}^1, \mathbf{x}^2, \ldots, \mathbf{x}^L \) are concatenated, along with the conditions $\mathbf{C}_s, \mathbf{C}_e$ and the condition feature vector \( \mathbf{C}_h \). 
This combined representation is passed through a 4-layer MLP to produce the final predicted displacement \( \mathbf{u} \in \mathbb{R}^{N \times 3} \). 
The predicted displacement is then applied to the input points to obtain the deformed point cloud \( \hat{\mathbf{y}} = \mathbf{u} + \mathbf{x^0} \).

\subsubsection{Force Prediction}
The force prediction branch takes as input the predicted displaced points \( \hat{\mathbf{y}} = \mathbf{u} + \mathbf{x^0} \), the initial point features \( \mathbf{h}^0\), the probe end position \( \mathbf{C}_e \), and the condition feature vector \( \mathbf{C}_h\) sampled along the probe line.

We first compute per-point difference vectors \( \hat{\mathbf{y}} - \mathbf{C}_e \) (broadcasting \( \mathbf{C}_e \)) and \( \tilde{\mathbf{h}}^0 - \mathbf{C}_h \), where \( \tilde{\mathbf{h}}^0 \) is the concatenation of the initial features \( \mathbf{h}^0 \) with the corresponding point \( x \) and \( y \) coordinates to align dimensionality with \( \mathbf{C}_h \).
These difference vectors are concatenated and passed through an MLP \( \phi_{\text{feat}} \), producing per-point latent features:
\begin{equation}
\mathbf{z}_i = \phi_{\text{feat}}\big( [\,\hat{\mathbf{y}}_i - \mathbf{C}_e;\, \tilde{\mathbf{h}}^0_i - \mathbf{C}_h\,] \big), \quad i=1,\ldots,N.
\end{equation}

We then aggregate the per-point features \( \mathbf{z} \) using global max and mean pooling, concatenate the pooled outputs, and apply a second MLP \( \phi_{\text{force}} \) to regress the scalar force \( F \in \mathbb{R} \):
\begin{equation}
F = \phi_{\text{force}}\big( [\, \max (\mathbf{z}); \; \mathrm{mean} (\mathbf{z})\,] \big).
\end{equation}

This design fuses spatial displacement and tissue feature differences relative to the probe interaction region, enabling the model to predict the applied force from localized and global deformation cues.

\section{Experiments}\label{experiments}

\begin{figure}
    \centering
    \subfloat[]{\includegraphics[width=0.4\textwidth]{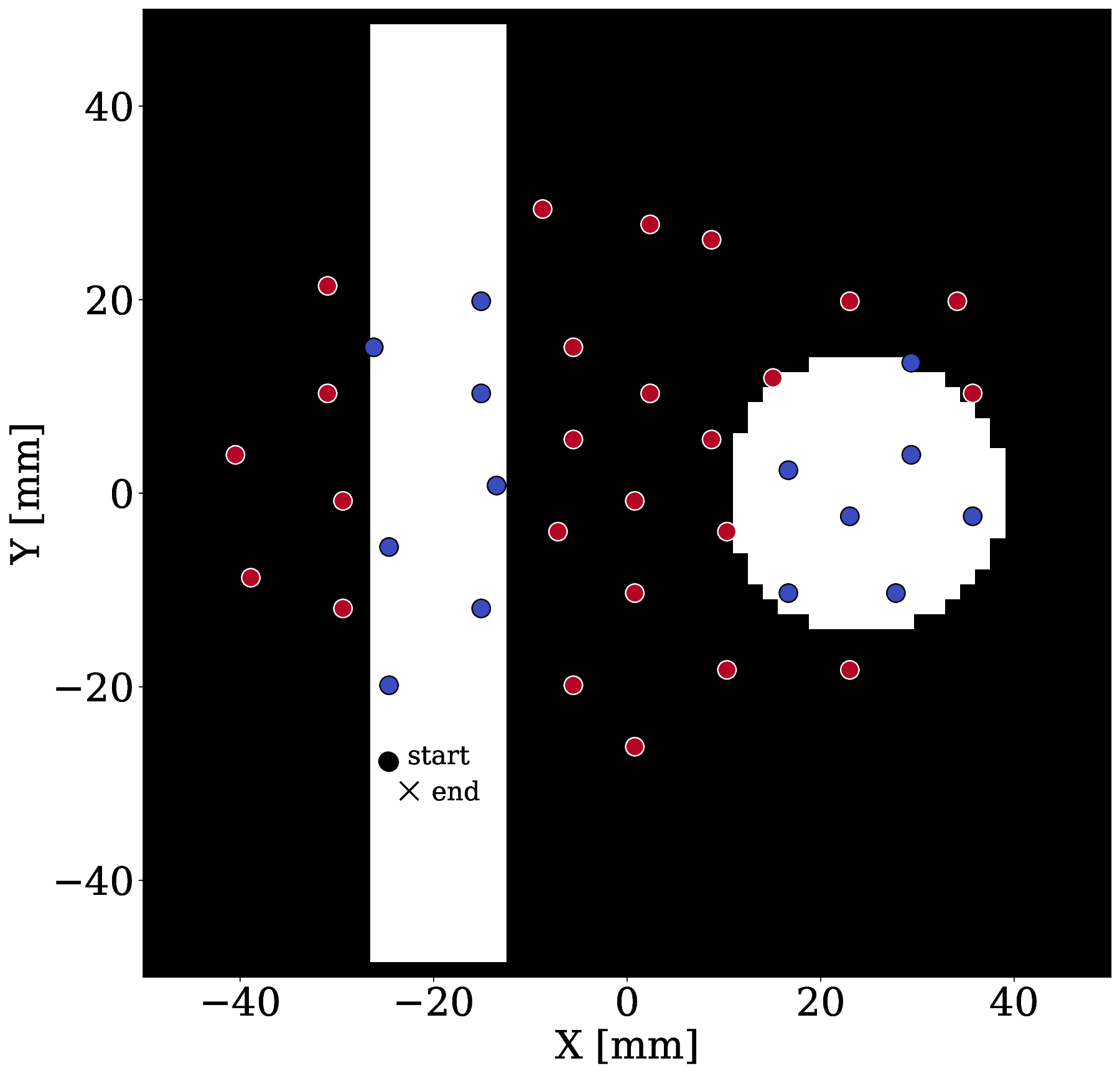}\label{exp_sample}}
    \subfloat[]{\includegraphics[width=0.4\textwidth]{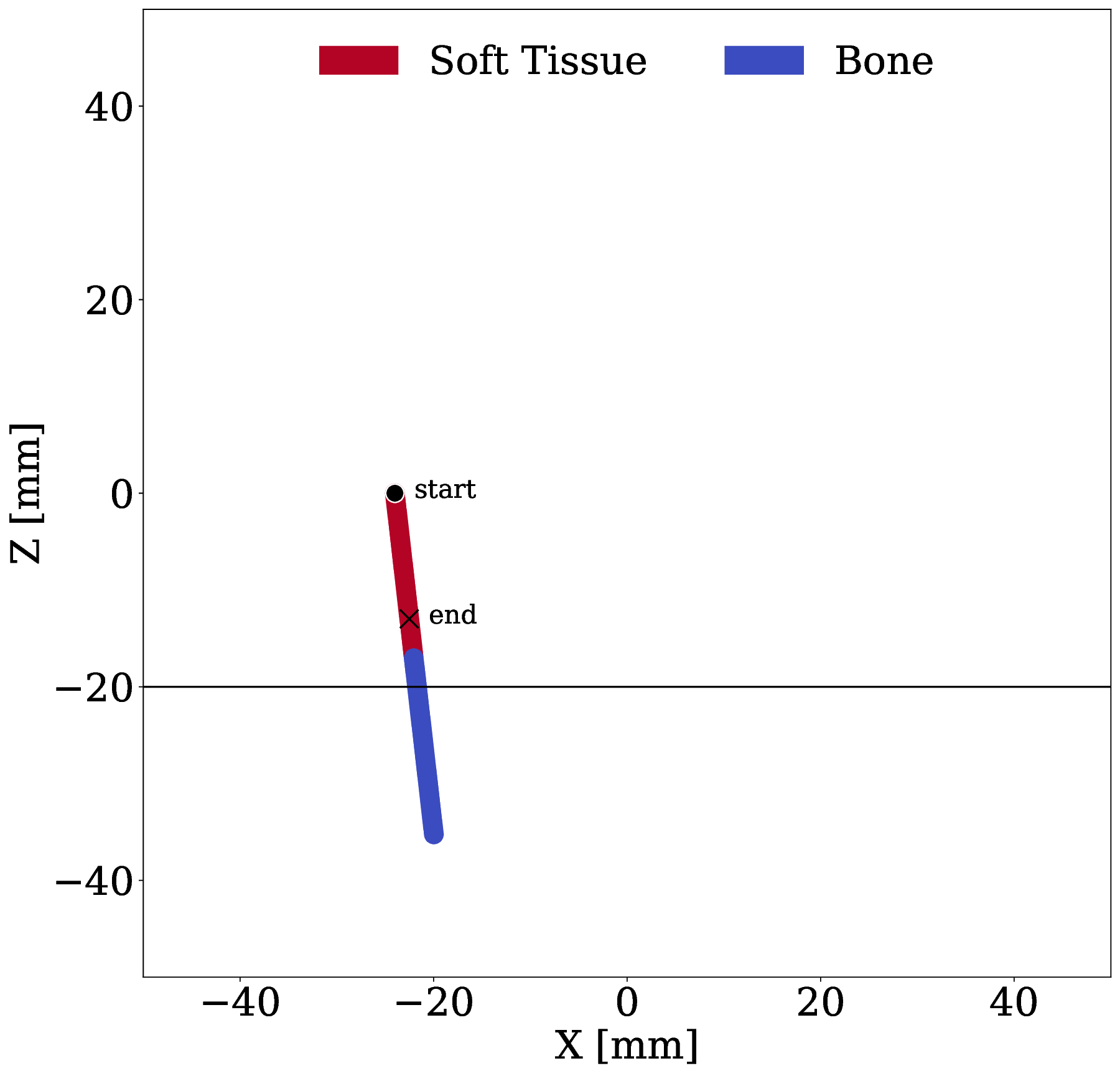}\label{cond}}
    \caption{\textbf{(a)} Top view of the input points from the FEM dataset sampled at experimental surface point locations, colored by their tissue features at a z-slice. \textbf{(b)} Tissue features of the condition with start and end indicating the location of the applied force.} 
\end{figure}

We conducted experiments on both simulated (FEM) and real (experimental) datasets to evaluate our model's performance under various conditions, including rotational generalization, point density change, and transfer learning for the experimental dataset. For comparison, we used the cGNN model \cite{cGNN2025} as a baseline. The baseline model is trained without the binary tissue features, only using the input point coordinates and the condition of applied force as an input. 

\subsection{FEM Dataset}
For training, we sampled $32 \times 32$ surface points on a regular grid from the FEM simulations.  
Each point is associated with binary tissue features as described in Section~\ref{sec:methods_tissue}. 
To ensure robust generalization, we split the FEM dataset into training, validation, and test sets with a ratio of 7:2:1, stratified by the deformation location. 
All training was performed using 5-fold cross-validation. 
To evaluate the generalization of the models, we additionally tested both our and the baseline model on rotated versions of the test set. 
Rotations were applied in $90^\circ$ increments around the z-axis.

\subsection{Experimental Dataset and Transfer Learning}
To evaluate performance on real-world data, we fine-tuned the FEM-trained models using the experimental dataset consisting of 40 measured surface points.
To ensure compatibility with this sparse input structure of the experimental data, the models were pre-trained on subsampled FEM data to the same 40 surface points as in the experimental data (see Fig.~\ref{exp_sample}).

Fine-tuning on the experimental dataset was performed using leave-one-out cross-validation, where each location was held out once for testing while the model is trained on the remaining ones. 
This strategy was chosen due to the small number of available deformation locations ($39$ in total), allowing us to make the most of limited data while ensuring a robust evaluation.
Furthermore, to test generalization beyond sparse settings, we evaluated these FEM-trained models on their full-resolution $32 \times 32$ point grid as well. 
The features of the datasets are shown in Table~\ref{features}.

\begin{table}
\centering
\setlength{\tabcolsep}{0.5em}
\caption{Features of the datasets with various input points, where $F_{max}$ is the maximum force for each indentation location, $d_{mean}$ is the average point distances across the dataset, and $tip_{mean}$ is the average tip distance between each data element. FEM (40) is the FEM dataset sampled with 40 points in the same location as the experimental surface points. FEM (1024) is the full FEM dataset sampled in a regular grid.}\label{features}
\begin{tabular}{l|ccc}
    \makecell[tl]{Dataset (points)} 
    & \makecell{$F_{max}$ [N]} 
    & \makecell{$d_{mean}$ [mm]} 
    & \makecell{$tip_{mean}$ [mm]} \\
    \midrule
    Experimental (40) 
    & $3.762\pm1.828$ & $1.512\pm3.898$ & $21.556\pm7.579$ \\ 
    FEM (40) 
    & $4.391\pm2.941$ & $1.548\pm3.674$ & $16.693\pm5.940$ \\ 
    FEM (1024) 
    & $4.391\pm2.941$ & $0.756\pm2.156$ & $16.693\pm5.940$ \\ 
\end{tabular}
\end{table}

\subsection{Loss Function}
The model was optimized using a weighted combination of displacement and force prediction losses. 
Specifically, we minimized a weighted displacement error $\mathcal{L}_d$ and a force prediction error $\mathcal{L}_f$ using mean squared error. The total loss was
$\mathcal{L} = \alpha \mathcal{L}_d + \beta \mathcal{L}_f$, 
where $\alpha = 1000$ and $\beta = 100$ were chosen to balance the magnitudes of the two loss terms.

To encourage more accurate predictions in regions of significant deformation, we introduced a \textit{relative weighting scheme} for the displacement loss. 
For each point, we compute its ground-truth deformation magnitude and use it to weight the corresponding loss term. 
Let $x_i$ be the original point, $y_i$ the ground-truth deformed point, and $\hat{y}_i$ the predicted deformed point. 
The displacement loss is defined as:

\begin{equation}
\mathcal{L}_d = \frac{1}{N} \sum_{i=1}^{N} w_i \cdot \left\| \hat{y}_i - y_i \right\|_2
\quad \text{with} \quad
w_i = \frac{\left\| y_i - x_i \right\|_2}{\max_i \left\| y_i - x_i \right\|_2 + \varepsilon}
\end{equation}

This weighting increases the penalty on errors at points where the ground-truth displacement is large (e.g., near the contact region), while de-emphasizing low-deformation areas. 
As a result, the model is encouraged to focus its learning capacity on regions most relevant for accurate deformation prediction. 
During evaluation, all reported deformation errors were computed using standard (unweighted) Euclidean distances, not the weighted loss used during training.

\subsection{Training Setup}
Models were trained for $200$ epochs using the Adam optimizer. The batch size was set to $64$ for the FEM dataset and $16$ for the experimental dataset. 
The learning rate was initialized to $10^{-3}$ for training on the FEM dataset and reduced to $10^{-4}$ for fine-tuning on the experimental dataset.

Each model used $k=5$ nearest neighbors for graph construction. 
Our proposed model architecture consisted of 4 stacked cEGCL layers. 
For $\phi_e$ and $\phi_x$, we used the same MLP construction as in \cite{ecgnn}.
Similarly, the additional $\phi_{cs}$ and $\phi_{ce}$ also consist of two-layer MLPs with one Swish activation function: $Input \rightarrow \{Linear \rightarrow Swish \rightarrow Linear\} \rightarrow Output$.

The MLP for the final prediction of displacement was implemented as four fully connected layers, with ReLU activations and dropout except for the final layer: $Input \rightarrow \{ Linear \rightarrow ReLU \rightarrow Dropout(0.3) \} \rightarrow 512 \rightarrow 256 \rightarrow 128 \rightarrow Output$.

The force prediction branch including the two MLPs $\phi_{\text{feat}}$ and $\phi_{\text{force}}$ were implemented as: 
\begin{itemize}
    \item The feature encoder $\phi_{\text{feat}}$: 
    $Input \rightarrow \{Linear \rightarrow ReLU\} \rightarrow 128 \rightarrow \{Linear \rightarrow ReLU\} \rightarrow 128$.
    \item The regressor $\phi_{\text{force}}$: 
    $128*2 \rightarrow \{Linear \rightarrow ReLU\} \rightarrow 512 \rightarrow \{Linear \rightarrow ReLU \rightarrow Dropout(0.2)\} \rightarrow 256 \rightarrow \{Linear \rightarrow ReLU \rightarrow Dropout(0.2)\} \rightarrow 128 \rightarrow Linear \rightarrow 1$.
\end{itemize}

All experiments were implemented in PyTorch using the PyTorch Geometric (v2.1.0) package and run on a single NVIDIA RTX 2080 GPU.

\section{Results}
Results in Table~\ref{tab_full} show that our model trained with tissue properties performs comparably to the baseline cGNN model, which does not incorporate tissue properties. 
However, under rotation, our model significantly outperforms the baseline. 
Our model also demonstrates a clear advantage in computational efficiency: a single simulation takes only \qty{0.015\pm0.001}{\s}, compared to \qty{0.056\pm0.002}{\s} for the baseline when tested on FEM data with 1024 points.

To assess the accuracy of depth prediction specifically, we also evaluated the relative tip error, defined as the absolute tip error divided by the ground truth tip distance.
Both models show low tip errors ranging from $2.3\%$ to $3.6\%$ in normal and rotated cases, for mean tip distances up to \qty{16.7}{\mm} within the dataset (Tab.~\ref{features}), with our model consistently achieving slightly lower errors.

The force prediction error is around \qty{0.6}{\N}--\qty{0.7}{\N} for both models, with our model showing improved robustness under rotation (\qty{1.2}{\N}) compared to the baseline (\qty{3.6}{\N}).

\begin{table}[b!]
\centering
\caption{Model performance on FEM dataset trained and tested on $1024$ points, with rot. indicating rotated test cases.}\label{tab_full}
\resizebox{\linewidth}{!}{%
\begin{tabular}{p{2cm}|ccccc}
    Model  
    & \makecell{Force\\Absolute\\Error [N]} 
    & \makecell{Mean\\Euclidean\\Distance [mm]}
    & \makecell{Max\\Euclidean\\Distance [mm]}
    & \makecell{Mean\\ Relative Tip\\Error [\%]}
    & \makecell{Time [s]}
    \\
    \midrule
    Baseline & \textbf{0.599$\pm$1.077} & \textbf{0.135$\pm$0.305} & $2.431\pm1.056$ & $3.1\pm2.7$
    & \multirow{2}{*}{$0.056\pm0.002$} \\
    Baseline \hfill rot. & $3.565\pm5.700$ & $0.316\pm0.614$ & $3.072\pm1.566$ & $3.6\pm3.3$ &
     \\
    \midrule
    Ours & $0.701\pm1.111$ & $0.156\pm0.298$ & \textbf{2.105$\pm$0.864} & \textbf{2.3$\pm$1.8}
    & \multirow{2}{*}{\textbf{0.015$\pm$0.001}} \\
    Ours \hfill rot. & \textbf{1.162$\pm$1.757} & \textbf{0.266$\pm$0.509} & \textbf{2.714$\pm$1.367} & \textbf{3.0$\pm$2.5}
    & \\
\end{tabular}
}
\end{table}

\begin{table}
\centering
\setlength{\tabcolsep}{0.4em}  
\caption{Model performance on sampled (40 points) and full (1024 points) FEM test set, and the experimental test set (Exp) with transfer learning.}\label{tab_sampled}
\resizebox{\linewidth}{!}{%
\begin{tabular}{p{1.2cm}r|cccc}
    Model & \makecell{Test set\\(points)}
    & \makecell{Force\\Absolut\\ Error [N]} 
    & \makecell{Mean\\Euclidean\\Distance [mm]}  
    & \makecell{Max\\Euclidean\\Distance [mm]}
    & \makecell{Mean\\ Relative Tip\\Error [\%]}
    \\
    \midrule
    Baseline & FEM (40) & \textbf{0.584$\pm$1.150} & \textbf{0.214$\pm$0.440} & $1.818\pm1.258$ & $2.0\pm2.5$\\
    Ours & FEM (40) & $0.697\pm1.022$ & $0.273\pm0.344$ & \textbf{1.366$\pm$0.751} & \textbf{1.4$\pm$1.2}\\
    \midrule
    Baseline & FEM (1024) & $1.867\pm2.009$ & $1.192\pm2.045$ & $8.199\pm2.866$ & $3.9\pm3.4$\\
    Ours & FEM (1024) & \textbf{1.080$\pm$1.504} & \textbf{0.257$\pm$0.377} & \textbf{2.564$\pm$1.078} & \textbf{1.9$\pm$1.5}\\
    \midrule
    Baseline & Exp (40) & \textbf{0.462$\pm$0.651} & $0.905\pm2.047$ & $9.307\pm5.354$ & $10.5\pm8.8$\\
    Ours & Exp (40) & $0.727\pm0.934$ & \textbf{0.901$\pm$1.810} & \textbf{7.824$\pm$5.196} & \textbf{3.2$\pm$3.3}\\
\end{tabular}
}
\end{table}

\begin{figure}
    \centering
    \includegraphics[width=0.9\textwidth]{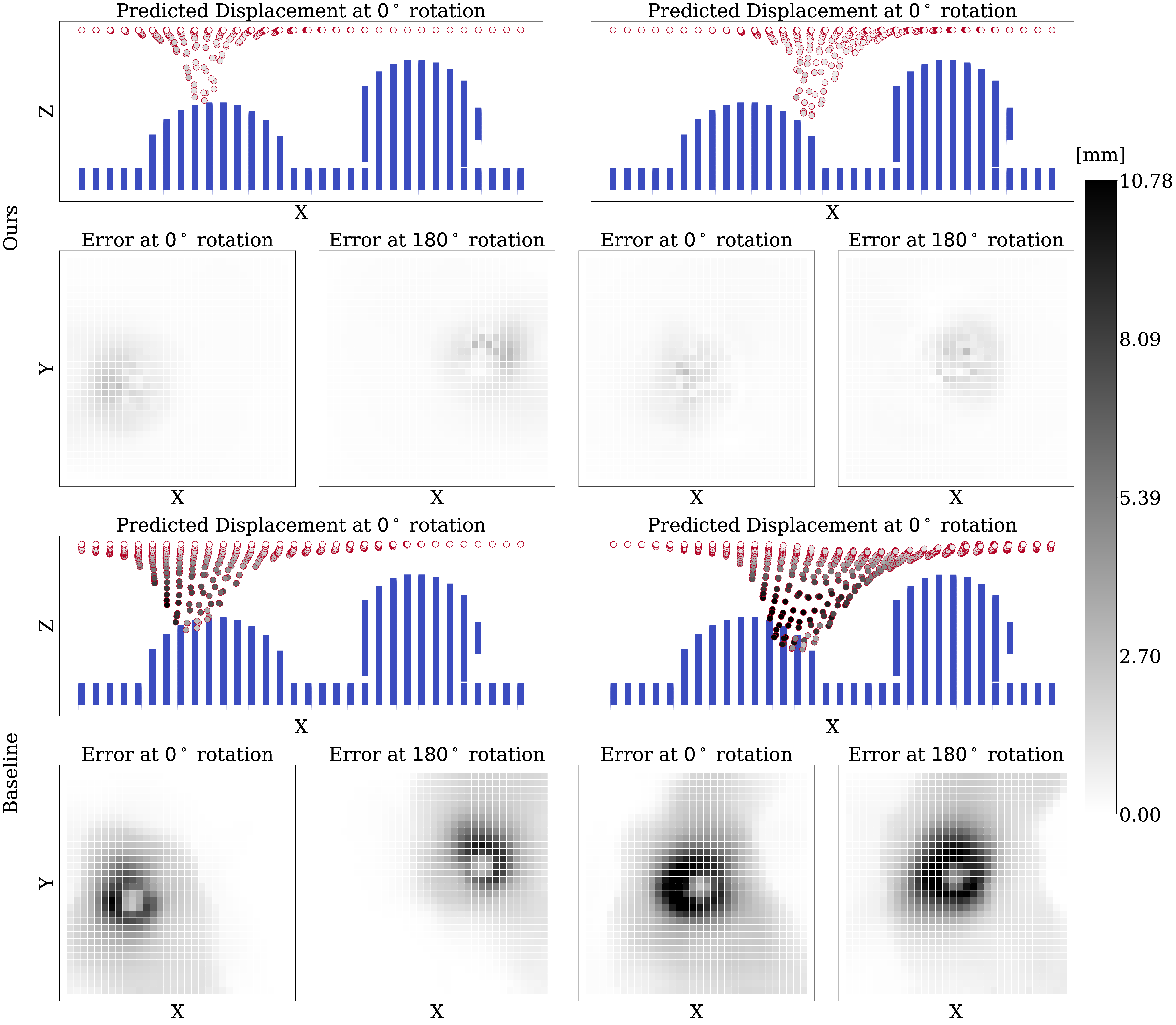}
    \caption{Displacement prediction results for FEM dataset trained on $40$ points, tested with $1024$ points, showing the distance errors at $0$° and $180$° rotation.}
    \label{fem_full}
\end{figure}

\textbf{Transfer Learning And Point Density} We evaluated the performance of both models after transfer learning on the experimental dataset (Tab.~\ref{tab_sampled}). 
In this setting, both models showed limited generalization: while the force prediction error remained similar as for FEM, the displacement errors were substantially higher, where both models reached a mean distance error of \qty{0.9}{\mm}. 
However, our model shows a lower maximum (\qty{7.8}{\mm}) and a relative tip error ($3.2\%$).

We also tested generalization across point densities. 
Table~\ref{tab_sampled} shows that when trained on sparse point clouds (40 points) and tested on denser configurations (1024 points), our model generalized successfully and outperformed the baseline, which failed in this setting (Fig.~\ref{fem_full}). 
This demonstrates robustness to sparsity and structure, a desirable property for real-world applications where dense, high-quality training data may not be available.

\section{Discussion}

The main limitation of this study is that our model's performance on the experimental dataset was significantly lower than on FEM simulations. 
This is likely due to a combination of factors: the experimental setup involved a limited number of poke locations (only 39), which is insufficient to cover the high variability of a heterogeneous anatomical structure. 
Additionally, measurement noise and post-processing errors, particularly around the contact tip, likely introduced inconsistencies. 
The force sensor used was also highly sensitive when contacting hard structures like bone, leading to unstable recordings, some of which had to be discarded. 
These limitations highlight the need for more robust and better-distributed experimental data to fully leverage the model's generalization capabilities.
However, the sub-millimeter prediction errors may still be within acceptable bounds for surgical simulation applications, where tolerances under \qty{1}{\milli\meter} are often considered clinically viable \cite{Olejnik2024,Viglialoro2019}.

Second, although we attempted to replicate the experimental scenario through FEM simulations, better-matching simulation setups may be necessary to bridge the gap between synthetic and real data. 
In particular, ensuring consistent material models and contact behavior could help reduce domain mismatch.

Despite this, our model showed strong generalization in rotated configurations and outperformed the baseline, indicating that combining equivariant convolutions with point-wise tissue property input improves robustness to changes in orientation. 
This suggests that the model may require fewer training examples under different orientations, reducing the need for extensive data augmentation or repeated retraining for each anatomical configuration, assuming similar material behavior across cases. 
Future work should confirm this by testing on a wider range of shapes and orientations.

Additionally, our model demonstrates adaptability to varying point densities and structures. When trained on sparsely sampled point clouds and tested on denser, unseen configurations, it successfully transferred learned deformations and produced low prediction errors. 
This flexibility reduces the need for retraining when input resolutions change and suggests that the model can operate effectively across different acquisition settings without requiring a fixed point count.

While our model predicts interaction forces alongside tissue displacement, its force prediction performance, especially under varied configurations including rotations, demonstrates promising applicability in surgical simulation contexts. 
Studies have shown that in real-time simulation and haptic rendering, absolute force accuracy, though important, is often secondary to the smoothness, stability, and responsiveness of feedback, which are critical for realistic tactile perception \cite{Okamura2009,Kuchenbecker2006}.
Consequently, approximate force feedback may still effectively convey tactile cues in surgical training simulators or planning tools \cite{Panait2009,Coles2011}. 
Additionally, our current approach treats each poke frame independently without temporal modeling. 
Incorporating temporal context using sequential models such as recurrent neural networks or attention mechanisms might improve force estimation by exploiting temporal consistency and dynamics inherent in surgical tool-tissue interactions. 
Such modeling would better reflect real surgical manipulation, where forces and deformations evolve smoothly over time, potentially enhancing both accuracy and realism in force feedback.

Finally, while we used segmented binary tissue classes (soft vs. bone) as input features, real-world applications may benefit from feeding raw CT or other medical imaging values directly into the model. 
This could eliminate the need for manual segmentation, which is time-consuming and error-prone, and enable seamless deployment on patient-specific scans.

\section{Conclusions}
We presented a graph neural network model that predicts soft tissue deformations and interaction forces by incorporating anatomical tissue properties with E(n)-equivariant message passing. 
Our approach achieves comparable or better accuracy than the baseline while being more robust to rotations and varying point cloud structures. 
Notably, it also demonstrates a significant speed-up, bringing us a step closer to enabling real-time surgical simulation and feedback.

The model generalizes well from sparse training data to denser test configurations, outperforming the baseline in such scenarios. 
Although performance on experimental data shows higher deformation errors, likely due to limited coverage and measurement noise, the deformation errors stay within sub-millimeter ranges. 
These findings support data-driven models as efficient and practical alternatives to traditional FEM simulations.

\section*{Financial disclosure}
This work was financially supported by the Werner-Siemens Foundation through the MIRACLE project.
%
%
%

\end{document}